\documentclass[11pt]{article}
\usepackage{fleqn,cospar}

\usepackage{url}

\usepackage{graphicx}
\usepackage[figuresright]{rotating}

\title{Models of Quasi-Periodic Variability in Neutron Stars and Black Holes}

\author{D.\ Psaltis\address{Center for Space Research,
Massachusetts Institute of Technology, Cambridge, MA 02139, USA}}

\begin{document}

% typeset front matter
\maketitle

\begin{abstract}
Black holes and weakly-magnetic accreting neutron stars show strong
quasi-periodic variability over timescales that span a very wide
range. This property offer us a unique opportunity to improve our
understanding of basic physical processes in strong gravitational
fields as it reveals, for the first time, phenomena that occur on the
most fundamental timescales near accreting compact objects.  In this
article, I review our current understanding of the variability
properties of accretion flows onto neutron stars and black holes. I
discuss a number of challenges faced by theoretical models, in order
to identify open questions in accretion physics that need to be
addressed. Finally, I discuss the relation to and implications for
variability models of the detection of nearly coherent oscillations
during type~I X-ray bursts in several neutron-star sources.
\end{abstract}

\section*{INTRODUCTION}

It is generally accepted that matter is accreted onto compact objects
at the high observed rates because of the turbulent transport of
angular momentum.  Since turbulent flows are typically variable, it is
not surprising that the X-ray brightness of even the so-called
steadily accreting objects is observed to show significant ($\sim
10-30$\%) fluctuations over a large range of timescales, from
milliseconds to weeks. 

The range of variability frequencies in an accretion flow is
determined by the extent of the flow itself, as well as by the
specific mechanism that produces them. As an example of the latter, in
a geometrically thin accretion disk, the motion of the gas is
primarily in the azimuthal direction, the flow of angular momentum is
in the radial direction, whereas the propagation of photons is in the
vertical direction. Depending on what causes the variability, the
accretion luminosity might, therefore, be modulated at any or all of
the three dynamical timescales that correspond to each of these
directions. All characteristic frequencies in an accretion flow depend
strongly on the distance from the central object. As a result, for a
process occurring throughout an accretion disk, one would expect strong
variability at all timescales, from milliseconds to days,
corresponding to the innermost and outermost parts of the accretion
disk, respectively. Indeed, the power spectra of accreting compact
objects show broad-band variability components at all these
timescales.  However, the most prominent and striking feature of their
power spectra is the presence of a small number of characteristic
frequencies, that appear in the form of narrow, quasi-periodic
oscillation (QPO) peaks.

Because of this observational fact, the aim of current models is the
identification of characteristic radii in the accretion flow that pick
only a small range of frequencies and hence produce narrow QPO peaks
in the power spectra. Such radii include the radius of the innermost
stable circular orbit (which is important in the diskoseismic models
of, e.g., Wagoner 2000), the magnetospheric radius (Alpar \& Shaham
1985), or the sonic-point radius (Miller, Lamb, \& Psaltis 1998).  The
simultaneous presence of multiple QPOs is then attributed to the
interaction of the flow with the stellar spin (as in, e.g., Miller et
al.\ 1998), to the presence of multiple characteristic radii in the
accretion flow (as discussed, e.g., in Miller et al.\ 1998), or to
different characteristic frequencies occurring at the same region in
the flow, as in the case of the diskoseismic models (see Wagoner 2000)
and the relativistic models of Stella et al.\ (1999) and Psaltis \&
Norman (2000).

Addressing which of all the characteristic frequencies or radii is
responsible for the observed variability properties of accreting
neutron stars and black holes is a subject of active research. In
particular, three important issues and their implications for
theoretical models are still a matter of debate: the importance of the
modulation of the accretion flow at the stellar spin frequency (as
required by beat-frequency models, in the case of accretion onto
neutron stars; see Miller et al.\ 1998), the suggested similarity
between the variability properties of neutron-stars and black holes
(Psaltis et al.\ 1999), and the relation of variability models with
the spin-frequency interpretation of nearly coherent oscillations
observed during type~I X-ray bursts (Strohmayer 2000, these
proceedings).

\subsection*{MODELS OF THE PERSISTENT VARIABILITY OF ACCRETING
COMPACT OBJECTS}

The motion of a fluid element in an accretion flow is determined by
the combined effect of several forces that act on it, such as gravity,
magnetic forces, radiation forces, and viscosity. The properties of
the different interactions allow us to define different timescales (or
frequencies), at which the various forces operate, and radii, at which
the effects of these interactions are dominant.  A number of
characteristic frequencies and radii that are often invoked in models
of the variability properties of compact objects are discussed below.
 
\subsection*{Characteristic Variability Frequencies}

A simple estimate of the three characteristic dynamical frequencies at
distance $r$ away from a compact object is given by the three
corresponding test-particle frequencies: the orbital frequency
in the azimuthal direction, the epicyclic frequency in the radial
direction, and the vertical oscillation frequency.  These correspond
to the characteristic frequencies of oscillation of the gas density in
a ring of non-interacting particles orbiting around a compact object,
when perturbed in the azimuthal, radial, and vertical directions,
respectively.  For a rotating compact object with an external Kerr
spacetime, they have three distinct and non-zero values, given by
(see, e.g., Perez et al.\ 1997)
\begin{equation}
   \Omega^2=\frac{M}{r^3[1\pm\alpha_*(M/r^3)^{1/2}]}\;,
\label{eq:omegak}
\end{equation}
\begin{equation}
   \kappa^2=\Omega^2[1-6M/r\pm8\alpha_*(M/r^3)^{1/2}-3\alpha_*^2/r^2]\;,
\label{eq:kappa}
\end{equation}
and
\begin{equation}
   \Omega_\perp^2=\Omega^2[1\mp 4\alpha_*(M/r^3)^{1/2}+3\alpha_*^2/r^2]\;,
\label{eq:omegap}
\end{equation}
respectively. Here, $M$ is the mass and $\alpha_*$ the specific
angular momentum per unit mass of the compact object, and the
fundamental constants are set to $c=G=1$. All three frequencies
describe periodic phenomena, which can reach a very high coherence when
confined to a narrow range of radii. Moreover, the azimuthal orbital
frequency~(\ref{eq:omegak}) is the highest of all characteristic
frequencies and is, therefore, expected to describe the fastest
phenomena occurring in the accretion flows. For these reasons, all
current models have invoked one or all of the above as the fundamental
frequencies of variability of accreting compact objects.

An accretion flow, however, is a hydrodynamic system and hence the
test-particle frequencies~(\ref{eq:omegak})--(\ref{eq:omegap}) can
only be approximations to the frequencies at which the flow is
modulated.  In the case of a geometrically thin accretion disk,
additional relevant timescales are (see, e.g., Frank et al.\ 1992),
the sound-crossing time given by
\begin{equation}
t_{\rm c} \equiv \frac{c_{\rm s}}{r}\simeq 
   \left(\frac{r}{h}\right)\frac{1}{f_{\rm K}}\;,
\end{equation}
where $c_{\rm s}$ is the sound speed and $h$ is the scale height of the
disk, and the viscous timescale, which is given by
\begin{equation}
t_{\rm V}=\frac{1}{\alpha_{\rm SS}} 
   \left(\frac{r}{h}\right)^2\frac{1}{f_{\rm K}}\;,
\end{equation}
where $\alpha_{\rm SS}$ is the viscosity parameter, and measures the
time it takes for a density inhomogeneity to be smoothed by
viscosity. The last two timescales do not describe periodic phenomena
and, therefore, are not expected to directly determine any
characteristic frequency of quasi-periodic variability.  However, the
effects they describe, such as the propagation of sound waves or
the viscous dissipation of density perturbations, can alter the
frequencies of oscillations in the accretion flows as well as
determine their lifetimes, and hence their coherence (see, e.g.,
Wagoner 2000; Psaltis 2000). Including such hydrodynamic effects is
both {\em necessary\/} for models to be physically self-consistent and
{\em required\/} by all current models for achieving agreement with
observations (see, e.g., Lamb \& Miller 2000; Psaltis 2000).

\subsection*{Characteristic Radii}

The observed QPOs in both neutron-star and black-hole systems are
typically narrow, with fractional widths as low as $\delta\nu/\nu\sim
10^{-2}$ (van der Klis 2000). Because all characteristic frequencies
in an accretion flow have a strong dependence on radius and height
above the equatorial plane, such small fractional widths severely
localize the physical mechanism that determines the QPO frequencies to
a narrow annulus in a geometrically thin component of the accretion
flow, so that $\delta r/r\le\delta\nu/\nu \sim 10^{-2}$ and $h/r\le
 (\delta\nu/\nu)^{1/2}\sim 10^{-1}$.

The notion of a narrow annulus in an accretion disk at a characteristic
radius, across which the fluid properties change considerably, is not
new in accretion theory and several such radii have been identified so
far.  For example, in the case of an accretion disk around a black
hole or a high-mass neutron star, the Keplerian flow is terminated at
the radius of the innermost stable circular orbit, inside which the
flow becomes quasi-radial and supersonic (see, however, Hawley \&
Krolik 2000). This radius depends only on the mass and spin of the
central object and, therefore, the dynamical frequencies that
correspond to it are very stable. This stability has been used both in
constructing models for QPOs in black-hole systems with frequencies
that show very little dependence on accretion rate (see, e.g., Wagoner
2000) and in placing model-independent upper bounds on the frequencies
of other variable-frequency QPOs in both black-hole and neutron-star
systems (see, e.g., Miller et al.\ 1998).

The presence of variable-frequency QPOs, however, in most neutron-star
and black-hole systems requires the presence of a characteristic
annulus with a radius that depends on accretion rate or any other
variable parameter of the system. For example, in the case of a
neutron star with a dynamically important magnetic field, the
Keplerian flow can be terminated at the so-called magnetospheric
radius at which magnetic torques can remove the angular momentum of
the accreting gas in a narrow boundary layer and bring it to
corrotation with the star. The radius of the boundary layer depends on
the mass accretion rate through the inner disk (see, e.g., Ghosh \&
Lamb 1978) and can become comparable to the neutron-star radius for
weak magnetic fields and high accretion rates. This radius has been
used in the magnetospheric beat-frequency models of the $\sim
20-60$~Hz QPOs observed from several neutron stars (see, e.g., Alpar
\& Shaham 1985).

The absence of any direct evidence of magnetic truncation of the
accretion disks around neutron stars that show high-frequency
variability as well as the existence of variable-frequency QPOs around
black holes necessitates the presence of characteristic radii that can
be generated and sustained without requiring a central object with a
hard surface or an anchored magnetic field. Radiation drag forces from
an external illuminating source can also remove efficiently the
angular momentum of the accreting gas, producing a sharp transition
from Keplerian to quasi-radial inflow (see, e.g., Miller et al.\
1998). This, so-called sonic point radius, can exist only very close
to the compact object ($\sim 3-10$ Schwarzschild radii) and is
invoked for setting the variability frequencies in the sonic-point
beat-frequency model of the high-frequency QPOs observed from neutron
stars. As another example, the transition in the thermal properties of an
unstable accretion disk during a dwarf-nova like instability cycle
is very sharp and has been used in models of nearly-coherent
variable-frequency oscillations observed from white dwarf systems (see
discussion below).

In summary, the requirements for a geometrically thin accretion disk
with a sharp transition in its properties are rather model
independent. They arise from three properties of the QPOs observed in
both neutron-star and black-hole systems: the fact that their
frequencies are highly variable, while, at the same time, they are
relatively coherent and strictly reproducible with spectral
state. These requirements challenge our current understanding of
accretion flows around compact objects.  The nature of variable
characteristic radii is not well understood and requires hydrodynamic
models of the accretion disks that take into account external forces
besides gravity. Moreover, the requirement for a geometrically thin
accretion disk close to a compact object accreting at near-Eddington
accretion rates (as in the case of the luminous source Sco~X-1, which
shows the most coherent oscillations at the highest inferred accretion
rates) requires the construction of accretion disk models at high
accretion rates that are both geometrically thin and stable, in
contrast to our current understanding (see, however, Abramowicz 1985).

\subsection{kHz QPOs: A signature of Spin-Orbit Interaction or of
Relativistic Effects?}

The most thoroughly studied of the variability phenomena observed
from an accreting compact object are the kHz QPOs discovered with the
{\em Rossi X-ray Timing Explorer\/} in the lightcurves of many
neutron-star systems. These are twin QPO peaks with variable
frequencies in the $\simeq 200-1200$~Hz range and a peak separation of
$\simeq 250-350$~Hz (see van der Klis 2000 for a review).

In the early observations, the peak separation of the two kHz QPOs
appeared to be constant and equal to the frequency of highly-coherent
oscillations observed during thermonuclear bursts from the same
sources (see below and Strohmayer 2000, these proceedings). This property
gave rise to the beat-frequency models of kHz QPOs (Strohmayer et al.\
1996; Miller et al.\ 1998) according to which the upper kHz QPO occurs
at the Keplerian frequency at a characteristic radius in the accretion
disk (the magnetospheric or the sonic-point radius) and the lower kHz
QPO occurs at the beat between the frequency of the upper kHz QPO and
the neutron-star spin frequency. Beat-frequency models rely on the
interaction between the orbital motion of the accreting gas and the
stellar spin through magnetic or radiation forces and can account, by
construction, for the similarity of the peak separation of kHz QPOs
with the frequencies of burst oscillations. However, they are
challenged by the subsequent discovery that the peak separation of the
kHz QPOs in all sources strongly decreases with increasing QPO
frequency (but see Lamb \& Miller 2000) and, moreover, by the
possibility that the same type of variability phenomena occur also in
black-hole systems, for which a beat-frequency model cannot be viable
(see Psaltis et al. 1999).

\begin{figure}[t]
\begin{minipage}{9.cm}
\includegraphics[width=65mm,angle=-90]{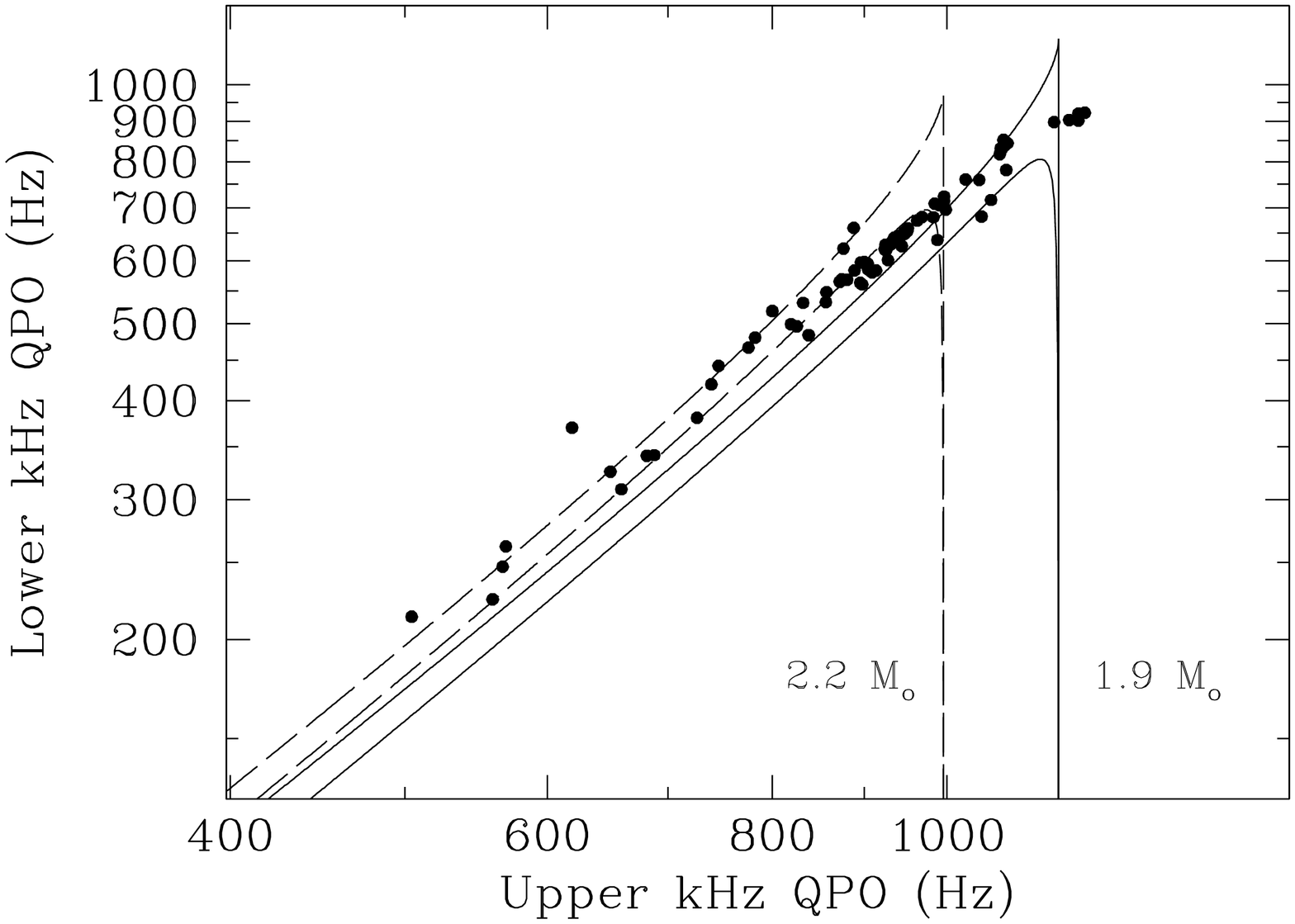}
\end{minipage}
\begin{minipage}{7.5cm}
\includegraphics[width=65mm,angle=-90]{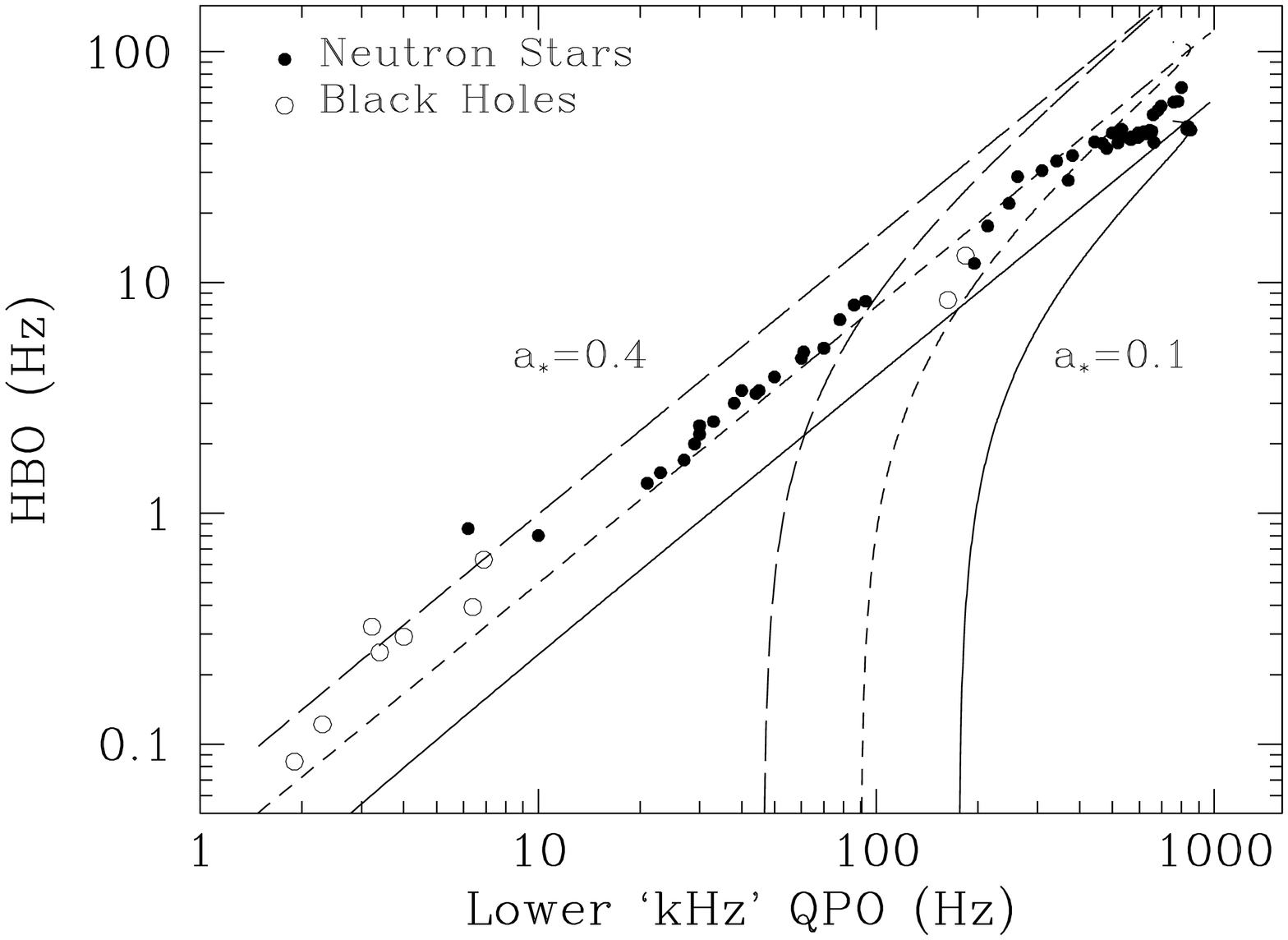}
\end{minipage}

\hfill

{\sf Fig. 1. Correlations of QPO frequencies in neutron-star and black-hole
systems. The curves are the predictions of the model in which all the QPO
frequencies are attributed to general relativistic frequencies in the
accretion flow (for details see Psaltis \& Norman 2000).}
\end{figure}

Detailed studies of three simultaneously detected QPOs in both
neutron-star and black-hole systems revealed that their frequencies
follow a small number of tight correlations that span up to three
orders of magnitude in frequency (Psaltis et al.\ 1999 and Fig.~1; in
the case of neutron stars, the three QPOs are the upper kHz QPO, the
lower kHz QPO, and a low-frequency, $\sim 20-60$~Hz, QPO previously
known). This property led to the construction of the so-called
relativistic models, in which the three QPO frequencies correspond to
the Keplerian, the periastron precession, and the nodal precession
frequencies of perturbed orbits in a characteristic radius in the
accretion disks (Stella, Vietri, \& Morsink 1999; Psaltis \& Norman
2000). Relativistic models can account for the presence of similar
variability phenomena in both neutron-star and black-hole systems and
their observed frequency correlations (see Fig.~1).  However, they do
not address directly the similarity of the peak separation of kHz QPOs
with the frequencies of burst oscillations.

An observable signature of each theoretical model is the spectrum of
characteristic frequencies it predicts beyond the ones currently
detected (see, e.g., Miller 2000b). For this reason, the inadequacy of
current data to distinguish between different models can be overcome
with the detection of additional QPO peaks. Even though some attempts
for detecting expected peaks at different frequencies have given
negative results (M\'endez \& van der Klis 2000a), a third kHz QPO has
been recently discovered in three low-luminosity sources (Jonker,
M\'endez, \& van der Klis 2000). Its properties were shown to favor
the relativistic model of kHz QPOs (Psaltis 2000), even though its
consistence with a beat-frequency model cannot be ruled out (see,
e.g., Alpar 1986; Jonker et al.\ 2000)

The fundamental difference, however, between the two categories of
theoretical models of neutron-star variability discussed above is the
role of the neutron-star spin frequency in determining some of the
observed QPO frequencies. None of the known kHz QPO sources has shown
so far pulsations at the neutron-star spin frequency in the persistent
emission. The only evidence for the magnitude of the spin frequencies
of sources that show persistent kHz QPOs comes from the modeling of
the highly coherent oscillations observed during thermonuclear bursts.
In the following section, I discuss the interpretation of burst
oscillations and its implications for models of the persistent
variability of neutron stars.

\section*{BURST OSCILLATIONS: INTERPRETATION AND IMPLICATIONS}

A large number of accreting neutron stars show recurrent type~I X-ray
bursts that are characterized by fast ($\sim 1$~s) rise and slower
exponential decay (Lewin et al.\ 1996) and are thought to be the
result of thermonuclear flashes on the surfaces of the neutron stars.
Their lightcurves are expected to be modulated at the stellar spin
frequencies because of the anticipated anisotropy of the thermonuclear
burning. Fast timing observations of accreting neutron stars during
thermonuclear bursts indeed showed highly coherent signals at $\sim
300-600$~Hz that have been interpreted as occurring at frequencies
that are within a percent of the stellar spin frequencies (see review
by Strohmayer 2000, these proceedings).

\section{Spin-Frequency Interpretation}

The spin-frequency interpretation of burst oscillations is motivated
mainly by two of their observed properties: their stability and
coherence. First, burst oscillations from the same neutron star source
have been shown to occur always at frequencies within a percent of each
other, even between bursts that were months apart. Second, when the
slow frequency evolution of burst oscillations that occurs at
timescales of seconds is removed, the coherence of the resulting
oscillations is large compared to any other QPO observed in the same
systems, by about an order of magnitude.  Given that the spin
frequency of a neutron star is highly stable and coherent, it is,
therefore, natural to assume that burst oscillations occur at the
stellar spin frequencies.

\begin{figure}[t]
\begin{minipage}{9.5cm}
\includegraphics[width=60mm,angle=-90]{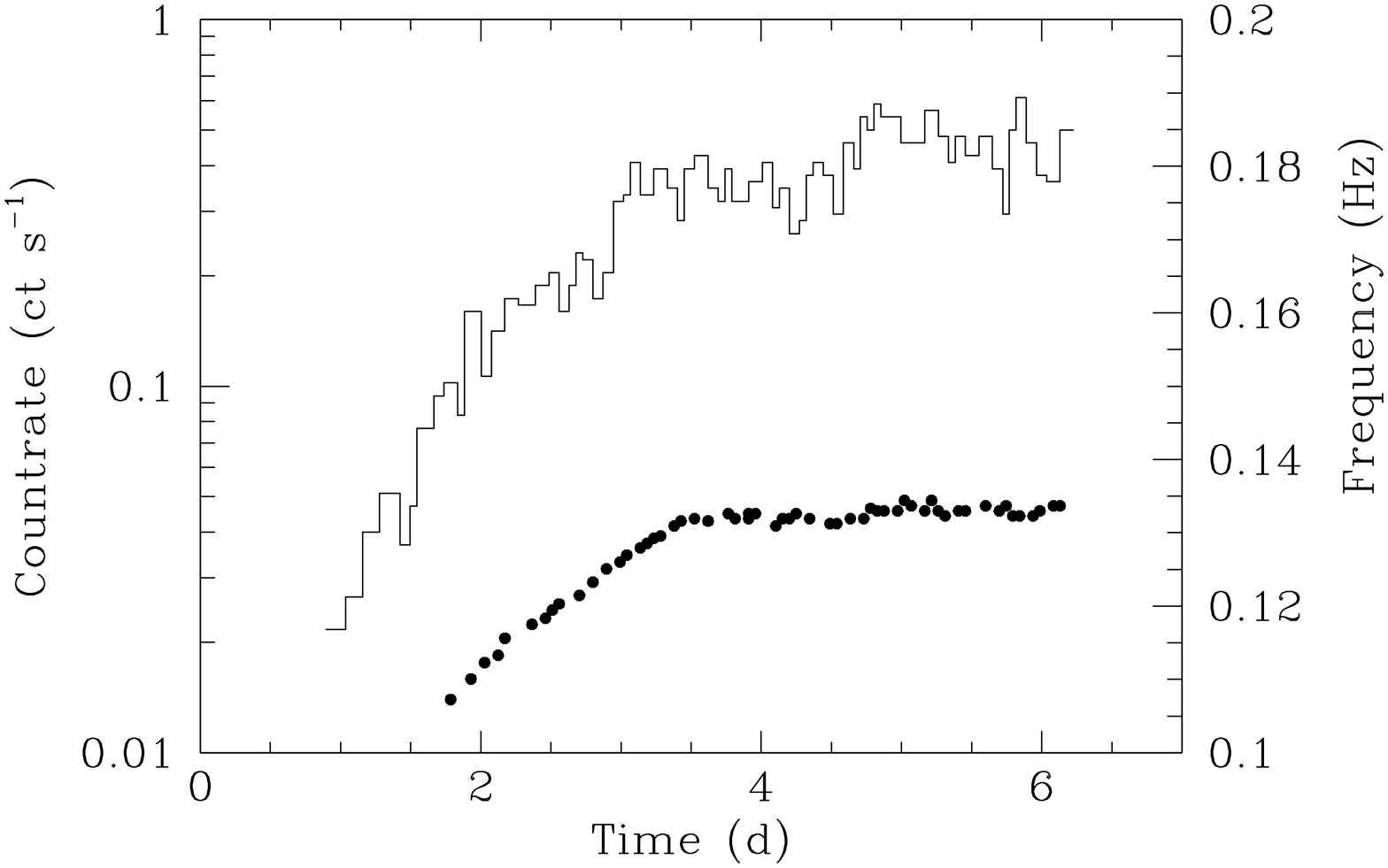}
\end{minipage}
\begin{minipage}{8.5cm}
\includegraphics[width=85mm]{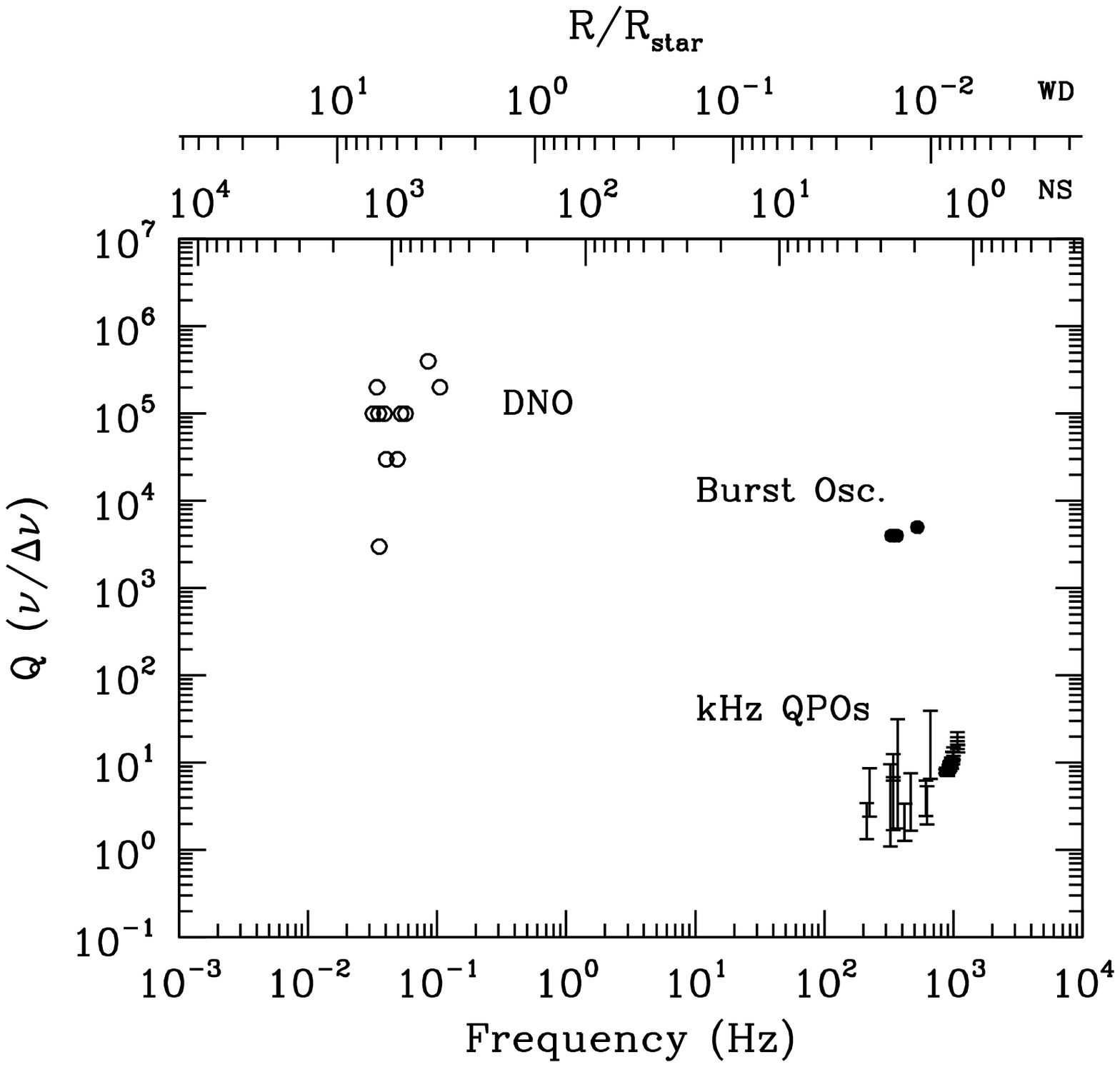}
\end{minipage}

\hfill

{\sf Fig. 2. {\em (Left)\/} Evolution of the EUV countrate and the
frequency of the dwarf-nova oscillations (DNO) in SS~Cyg (Mauche
1996). {\em (Right)\/} Coherence of various observed oscillations; the
kHz QPOs correspond to the Z sources, the burst oscillations to
KS~1731$-$260, 4U~1702$-$90, and 4U~1728$-$34, and the DNO have been
taken from the compilation of Patterson (1981). The top axis shows the
radius at which the orbital frequency is equal to the observed
frequency, for a $0.5 M_\odot$ white dwarf ($R_{\rm star}=10^8$~cm) and a
$2 M_\odot$ neutron star ($R_{\rm star}=10^6$~cm).}
\end{figure}

It is important to note, however, that the above arguments are neither
strong nor conclusive. For example, there exist several other
frequencies in the systems that are as stable as the stellar spin
frequency, such as the Keplerian frequency at the innermost stable
circular orbit, or the maximum radial epicyclic frequency. The
coherence ($Q\equiv \nu/\delta\nu\sim 10^3)$ of the burst oscillations
might also be higher than any other known QPO in neutron-star systems,
but is not comparable to the coherence of the periodic pulsations of
rotation- or accretion-powered pulsars (even though this might be just
due to the limits on the coherence imposed by the short lifetime of
the oscillations). Moreover, QPOs with significantly higher coherences
($Q\simeq 10^4-10^6$) have been detected in the high-energy emission
of several accreting white-dwarf systems (the so-called dwarf-nova
oscillations; see, e.g., Patterson 1981). The properties of dwarf-nova
oscillations are strikingly similar to those of burst oscillations
(see Fig.~2). Their frequencies are variable at timescales of days and
scale with inferred luminosity, so that they cannot be related to the
white-dwarf spin-frequencies but are probably phenomena in the
accretion flows. When their slow frequency evolution is removed, the
coherence of the oscillations can reach values as high as
$10^6$. Finally, their frequencies are comparable to the dynamical
timescale at a few white-dwarf radii, as is the case of
burst-oscillation frequencies in neutron stars (see Fig.~2). None of
these arguments suggests that dwarf-nova oscillations and burst
oscillations are produced by the same mechanism. They show, however,
that accretion disk phenomena can produce periodic oscillations of the
accretion flows very close to the central objects, with very
high coherences.

The spin-frequency interpretation of burst oscillations faces also a
number of additional challenges. For example, the stringent upper
limits on the amplitudes of any harmonics of the fundamental
frequencies strongly contradict model predictions. More importantly,
the detection of burst oscillations with amplitudes as high as 10\% at
the tails of the bursts requires that the burning front has ignited
only $\sim 30$\% of the neutron star surface, even several (10-20)
seconds after the start of the burst, which is hard to understand.
Finally, the interpretation of the frequency evolution of burst
oscillations in terms of a decoupling of the burning layer from the
rest of the star cannot account for the large (up to $1.3$\%; Galloway
et al.\ 2000) observed fractional change of the oscillation
frequencies. The interpretation of the frequency evolution of burst
oscillations becomes also more complicated, given the fact that in a
number of bursts, multiple oscillation peaks have been detected
simultaneously with comparable yet distinct frequencies (see, e.g.,
Miller 2000a; Galloway et al.\ 2000; see also Fig.~3).

\begin{figure}[t]
\begin{minipage}{10.cm}
\includegraphics[width=85mm,angle=0]{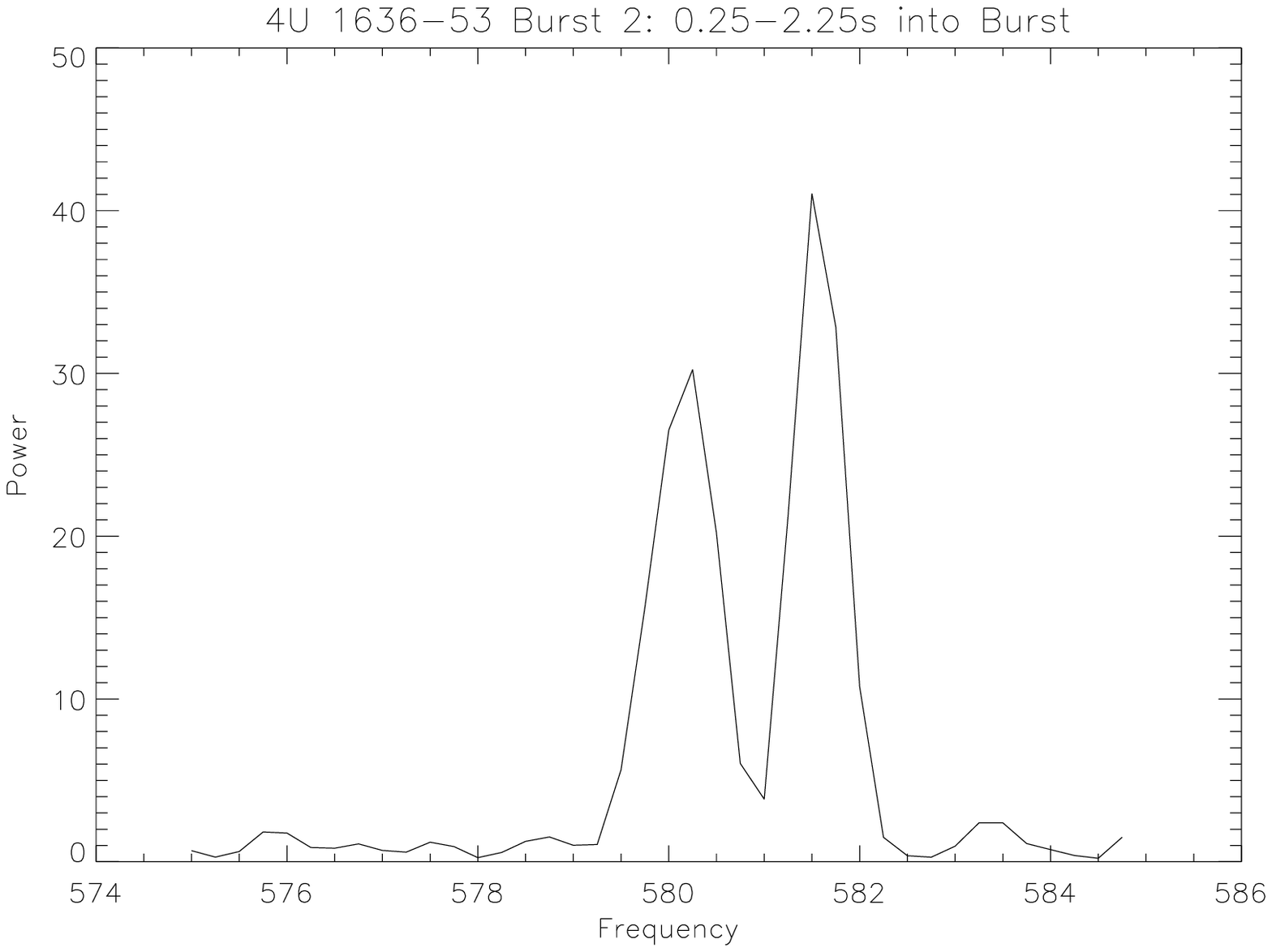}

{\sf Fig. 3. ({\em Left\/}) Detail of the power spectrum of the first
two seconds of a burst observed from the source 4U~1636$-$53, showing
the simultaneous presence of two distinct oscillation peaks; these
peaks are also seen simultaneously in the dynamical power spectra and
therefore cannot be due to the rapid evolution of one oscillation
frequency (see, e.g., Miller 2000a). ({\em Right\/}) Amplitudes of the
lower (filled symbols) and upper (open symbols) kHz QPOs in (a)
4U~1728$-$34 and (b) 4U~1608$-$52 as a function of their frequencies
(from M\'endez et al.\ 2000b).}
\end{minipage}
\begin{minipage}{8.cm}
\includegraphics[width=70mm,angle=0]{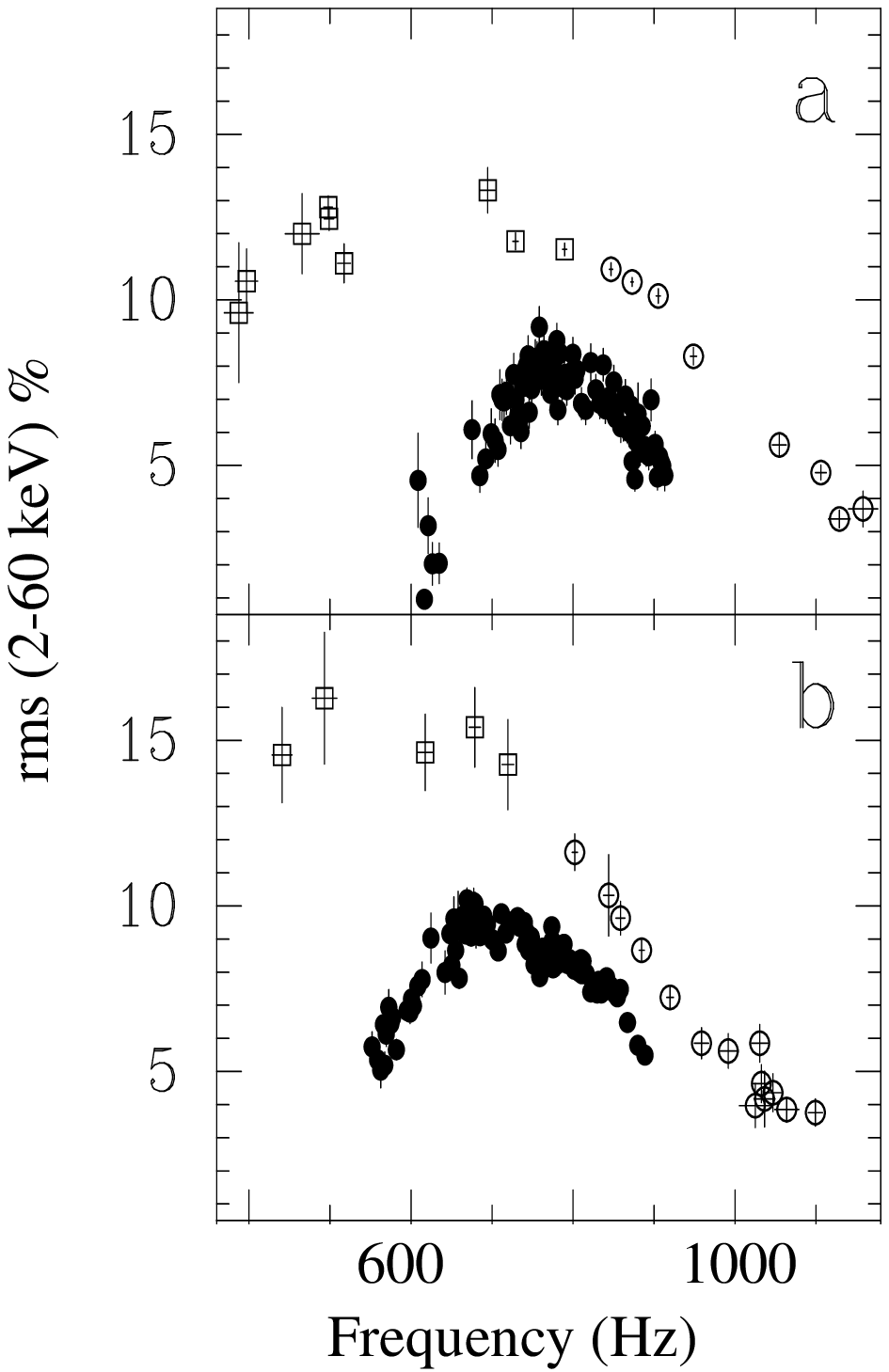}
\end{minipage}
\end{figure}

Despite all these issues, the detection of oscillations during type~I
X-ray bursts from the millisecond pulsar SAX~J1808.4$-$3086 at the known
spin-frequency (in't Zand et al.\ 2000) provides strong support to the
spin-frequency interpretation of burst oscillations.  It is possible
that the oscillations observed during type~I bursts from this source
are the anticipated signatures of non-uniform burning on the
neutron-star surface (given that the millisecond pulsar is the only
known, weakly-magnetic accreting neutron star with a non-uniform
surface emission during the persistent emission), whereas the burst
oscillations observed from other sources that show no pulsations in
their persistent emission is not. Conclusive identification of the
oscillations observed in SAX~J1808.4$-$3086 as bona fide burst
oscillations or detection of the neutron stars spin frequencies in the
persistent emission of sources that show burst oscillations will settle
the issues discussed above.

\section{Implications for Beat-Frequency Models}

The spin-frequency interpretation of burst oscillations provides the
motivation for and the strongest argument in favor of the
beat-frequency interpretation of the kHz QPOs in neutron-star
systems. However, in some sources burst oscillations are detected with
fractional amplitudes that reach $\sim 100$\% while the upper limits of
any coherent pulsations in the persistent emission of the same sources
are 2-3 orders of magnitude smaller. It is still puzzling how the
asymmetry during thermonuclear burning can produce such strong
oscillations at the stellar spin frequency (so that the compactness
and orientation of the system are not unfavorable), the asymmetry
during the persistent emission can produce a strong beat-frequency QPO
between a Keplerian frequency and the stellar spin (so that the
accretion flow interacts with the stellar spin), and yet the persistent
emission is not modulated at the stellar spin frequency.

In several sources, the maximum peak separation of the kHz QPOs is
comparable (to within several percent) to the frequency of burst
oscillations or to half of that frequency. This observational fact
together with the spin-frequency interpretation of the burst
oscillations require that either (a)~the peak separation of the kHz
QPOs is {\em always\/} comparable to the stellar spin frequency and
the burst oscillations occur at the spin frequency or its overtone or
(b)~the burst oscillations occur always at the stellar spin frequency
and the peak separation of the kHz QPOs is comparable to the spin
frequency or its {\em sub-}harmonic. Note that the above requirement
needs to be addressed by any model of the kHz QPOs; the beat-frequency
interpretation requires that option (a) is correct.

Identifying, however, the frequency of the $\sim 600$~Hz burst
oscillations with the first overtone of the stellar spin frequency, as
required by beat-frequency models, is hard to justify. Such an
identification would require, in some sources, the simultaneous
ignition of thermonuclear burning on two, almost exactly antipodal
spots on the stellar surface, presumably its magnetic poles. The
frequency evolution and coherence of such burst oscillations would
necessitate that the burning layers at the two antipodal spots
decouple from the rest of the star and are subject to the same
differential rotation. Finally, the absence of any oscillations with
comparable amplitudes at the postulated spin frequencies (i.e., at
half the observed frequencies) would require fine tuning in the
geometry of the systems (both ignition spots to lie on the rotational
equator and the observer to view the systems very nearly edge-on)
which has a very small a priori probability to occur and is,
therefore, inconsistent with the fact that the majority of burst
oscillation sources (six out of nine; see Muno et al.\ 2000) show
burst oscillations at twice the peak separation of the kHz QPOs.

\section{Implications for Relativistic Models}

In the relativistic models of kHz QPOs, the spin frequency of the
neutron star does not determine directly the frequencies of any of the
QPOs. As a result, the spin-frequency interpretation of the burst
oscillations does not have direct implications for such
models. However, it is uncomfortable, within a relativistic model, to
rely on pure coincidence in order to explain the observed similarity
between the burst oscillation frequencies and the peak separations of
the kHz QPOs or its harmonic. Moreover, the stellar spin frequency
determines the properties of the spacetime and hence the magnitudes
and correlations of the predicted QPO frequencies. As a result,
measurements of the neutron-star spin frequencies constrain
relativistic models in two ways.

First, the peak separation of the kHz QPOs, in the relativistic
models, is comparable to the radial epicyclic frequency at a
characteristic radius in the accretion disk. The spin-frequency
interpretation of the burst oscillations would therefore require
either (a) the spin frequencies of the neutron stars to evolve towards
the maximum radial epicyclic frequency in the accretion flow, which is
hard to justify, or (b) one or both of the kHz QPOs to attain their
high observed amplitudes when their peak separation is comparable to
the stellar spin frequency. The latter option is also motivated by the
fact that the amplitude of the lower kHz QPO in several sources is
observed to increase with frequency, reach a maximum, and decrease
again over a relatively narrow range of frequencies so that the QPO is
undetectable when the predicted peak separation is significantly
different than the burst oscillation frequency (see Fig.~3). Such a
possibility can be accommodated in the hydrodynamic relativistic model
of Psaltis \& Norman (2000), if the accretion flow at the
characteristic radius is also perturbed at the stellar spin frequency in
the radial direction, giving rise to a strong resonance only when the
radial epicyclic frequency is comparable to the stellar spin..

\begin{figure}[t]
\begin{minipage}{9.cm}
\includegraphics[width=65mm,angle=-90]{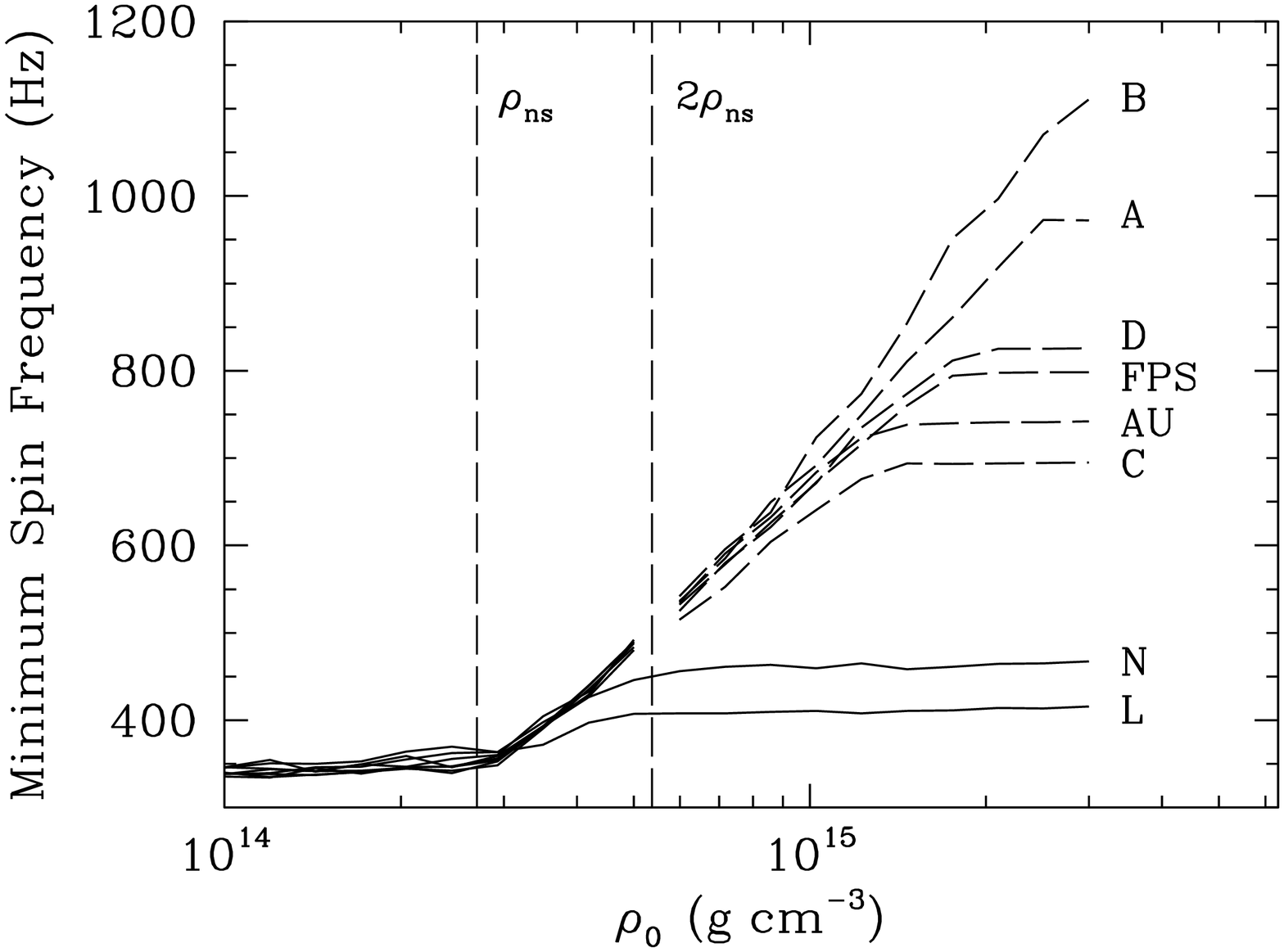}
\end{minipage}
\begin{minipage}{7.5cm}
\includegraphics[width=65mm,angle=-90]{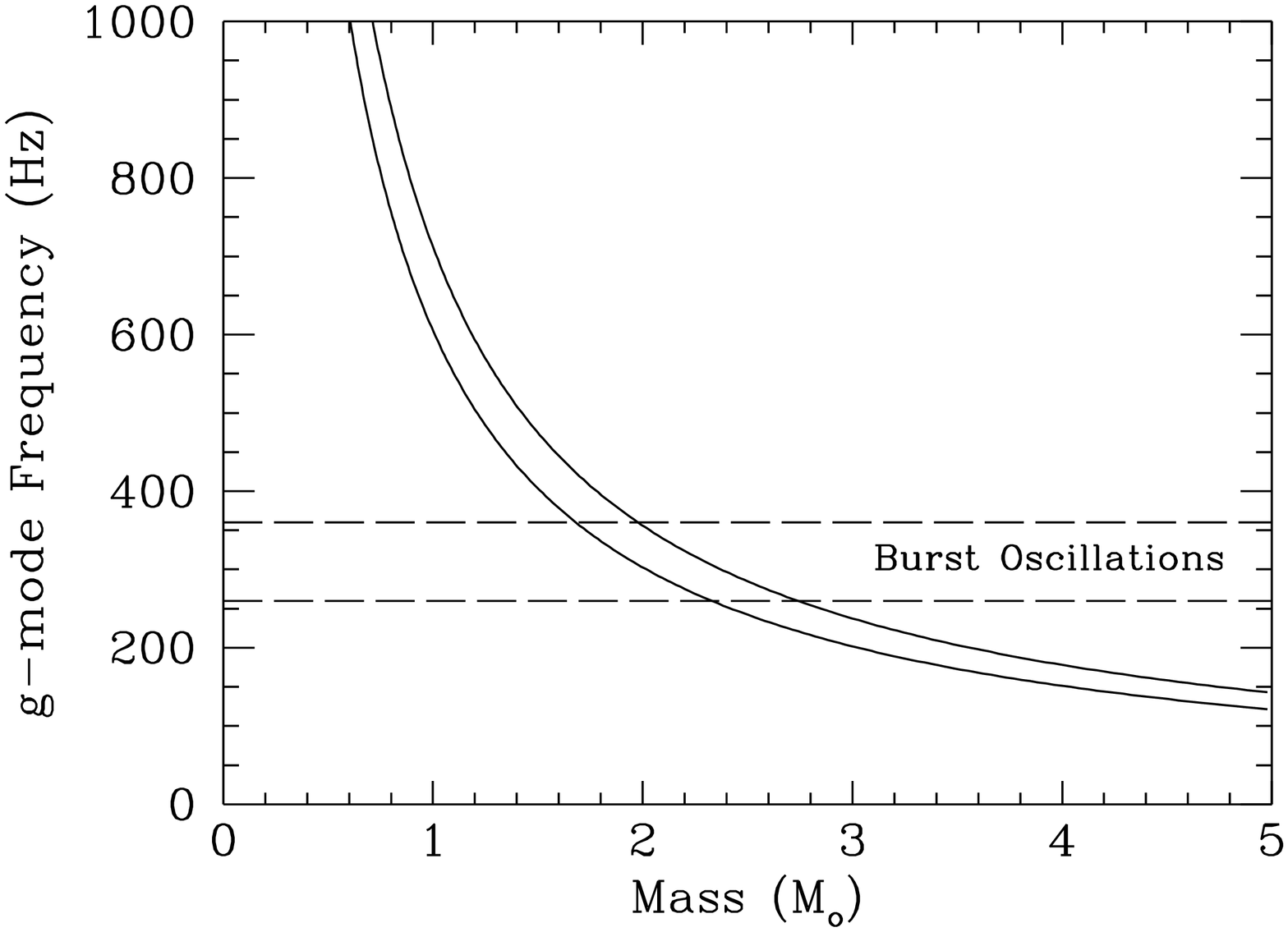}
\end{minipage}

\hfill

{\sf Fig. 4. {\em (Left)\/} Minimum neutron-star spin-frequency
required by the relativistic model of QPOs, for different
equations-of-state, if the observed low-frequency QPOs occur at the
first overtone of the Lense-Thirring frequency. {\em (Right)\/} Range
of predicted $g-$modes frequencies for a Kerr spacetime and different
hydrodynamic corrections (for details see Wagoner 1999).}
\end{figure}
 
Second, the frequency of the third ($\simeq 20-60$~Hz) QPO observed in
neutron-star systems and identified with the nodal precession
frequency in the relativistic models, is proportional to the spin
frequencies of the neutron stars and their moments of inertia. It can
be shown that for any equation of state of neutron-star matter, the
observed QPO frequencies are too high to be produced around a neutron
star that is spinning at the frequencies inferred from burst
oscillations (Psaltis et al.\ 1999). This is also shown in Fig.~4,
were different equations-of-state and densities (up to which a given
equation-of-state is assumed to be valid) are used in inferring the
minimum spin frequency required for the observed QPO frequencies to be
equal to twice the nodal-precession frequencies in sources such as
Sco~X-1 and 4U~1728$-$34.  In the latter source, the required
$>500$~Hz spin frequency is substantially higher than the $\sim
360$~Hz spin frequency inferred from burst oscillations.
 
\section{An Alternative}

The only theoretical idea that has been developed in any detail so far
for explaining the properties of the burst oscillations is that of a
burning surface layer that can decouple from the rest of the star and
rotate at a frequency comparable but not strictly equal to the
neutron-star spin frequency. It will be instructive, however, if other
possibilities are also explored and tested against observations. In fact,
the relativistic models of the persistent-emission QPOs can be used
as motivation to exploring models in which the burst oscillations are 
generated in the accretion flows and not on the stellar surfaces.

In the relativistic models, the peak separation of the kHz QPOs is
comparable to the epicyclic frequency in a characteristic radius in
the accretion flow. This frequency has a maximum at a radius very
close to that of the innermost stable circular orbit that depends only
on the mass and the spin of the compact object and is, therefore, very
stable. Moreover, near this maximum, a frequency cavity is produced in
which disk modes (the so-called $g$-modes) can be efficiently trapped
(see Wagoner 2000). The $g-$modes can be excited when the
thermonuclear flash disturbs the inner accretion flow and can last for
a time comparable to the viscous timescale, which is of order of
several seconds. The frequencies of $g-$ modes are comparable to the
maximum epicyclic frequency in the accretion flow, are equal to the
observed $\sim 300$~Hz frequencies for neutron stars with masses $\sim
2 M_\odot$, and alleviate the need for most known weakly-magnetic
neutron stars to be spinning at very similar frequencies. The
frequencies of the lower-order modes are comparable to the maximum
peak separation of the kHz QPOs, as predicted by the relativistic
models, whereas the frequency of the next higher-order mode is $\sim
4$ times larger and not harmonically related, which will account for
the absence of harmonic structure in the observed power spectra of
bursts. Moreover, theoretical calculations show that the oscillatory
spectra of $g-$modes contain several closely spaced modes (Wagoner
2000), which would appear in the observed power spectra as the
simultaneously detected peaks shown in Fig.~3. Finally, assuming that
the accretion (and not {\em only\/} the burst) luminosity is modulated
at the burst oscillation frequency, then the inferred fractional
amplitudes of burst oscillations will be significantly smaller than
currently reported.

It important, however, to stress that a disk-mode interpretation of
the burst oscillations faces a number of challenges, as well. For
example, even though accretion modes seem to be able to reach very
high coherences (as observed, e.g., in the dwarf-nova oscillations),
current studied of $g-$modes limit their coherence to values smaller
than observed. More importantly, though, the conclusive identification
of the oscillations observed in SAX~J1808.4$-$3086 as bona fide burst
oscillations will offer conclusive evidence against such
interpretations.

\section*{TESTING FUNDAMENTAL PHYSICS WITH COMPACT-OBJECT VARIABILITY}

Theoretical models of the variability properties of accreting neutron
stars and black holes are still in their infancy. Important questions,
such as the importance of the stellar spin or the steady-state
structure of geometrically thin accretion disks at high accretion
rates need to be addressed before the properties of time-dependent
accretion can be modeled in detail. However, even in their current
state, models of the variability of accreting neutron stars and black
holes show the potential of such studies in probing the equation of
state of neutron-star matter beyond the reach of laboratory
experiments and in searching general relativistic effects that occur
only in the strong-field regime.

\begin{figure}[t]
\begin{minipage}{9.cm}
\includegraphics[width=65mm,angle=-90]{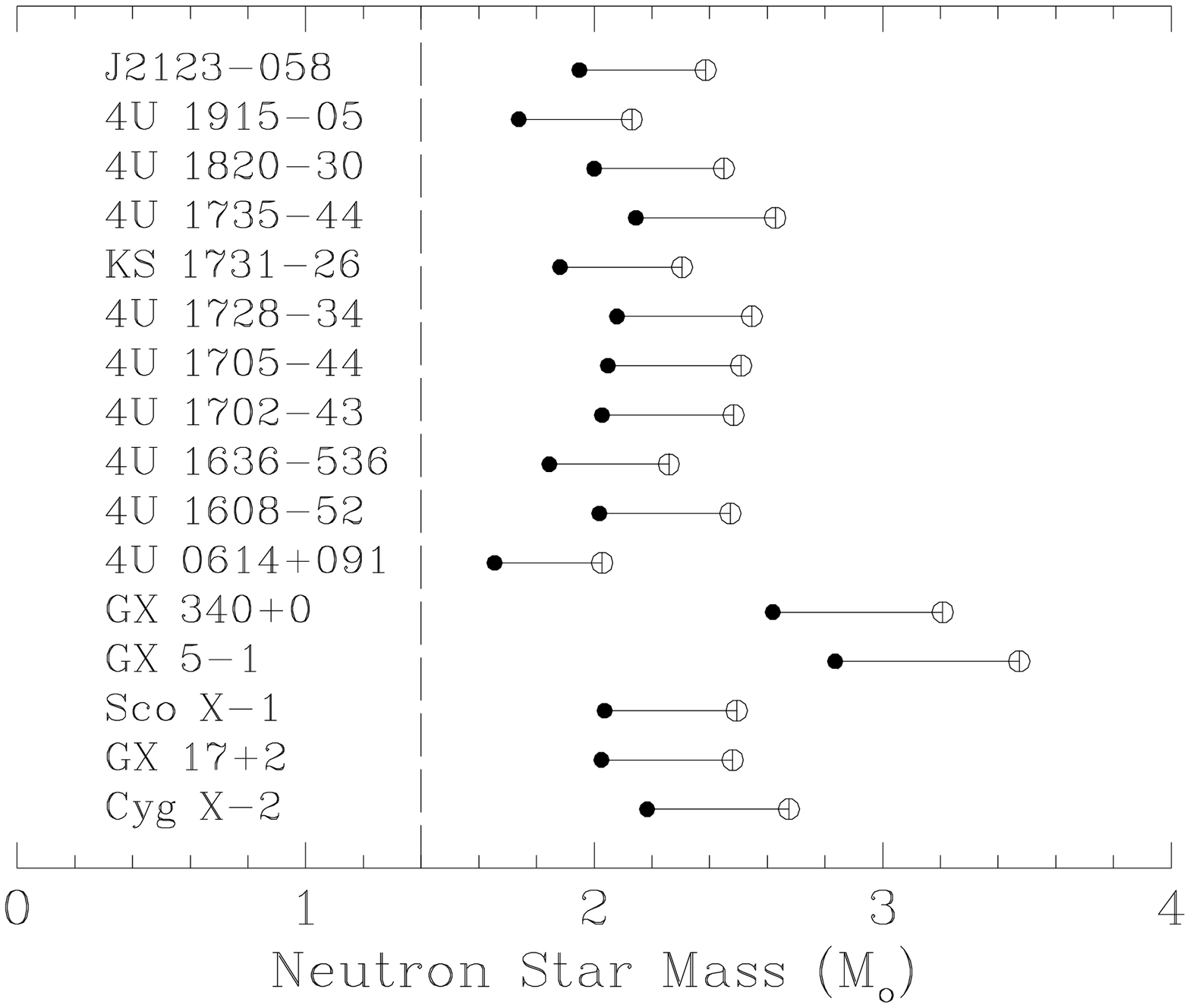}
\end{minipage}
\begin{minipage}{7.5cm}
\includegraphics[width=65mm,angle=-90]{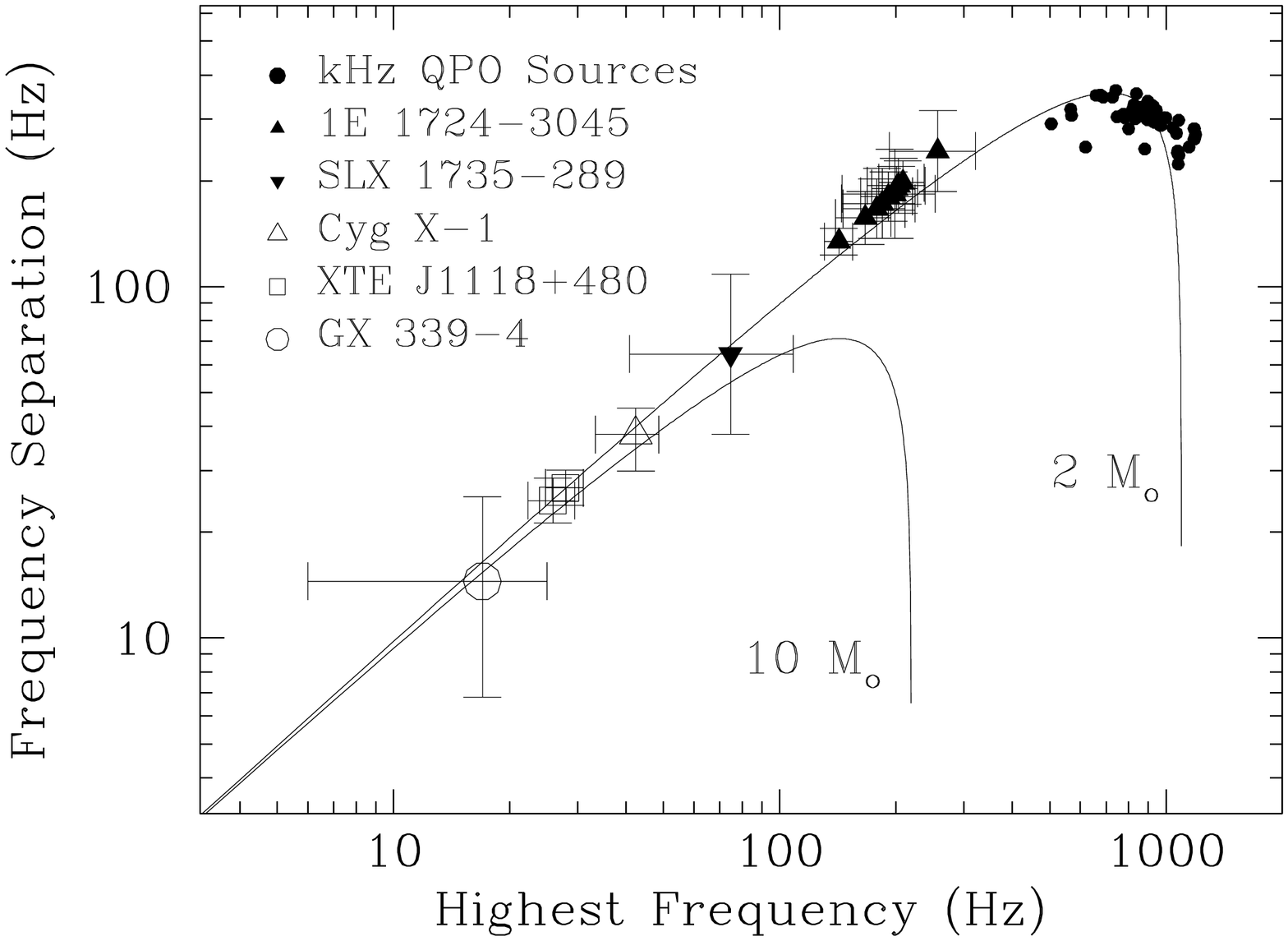}
\end{minipage}

\hfill

{\sf Fig. 5. {\em (Left)\/} Maximum neutron-star masses for individual
sources allowed by the current observations; filled and open circles
correspond to non- and slowly-rotating ($\alpha_*=0.3$) stars (after
Chakrabarty et al.\ 2000).  {\em (Right)\/} Frequency separation of
the highest observed frequencies in different neutron-star (filled
symbols) and black-hole (open symbols) sources as a function of the
highest observed frequency; the lines correspond to the predictions of
the relativistic model (after Belloni et al.\ 2000).}
\end{figure}

For a given neutron-star spin frequency, the equation of state
determines the gravitational mass and radius of the neutron star and
hence the Keplerian frequency at the innermost stable circular orbit.
The latter is the highest possible frequency of any long-lived,
quasi-coherent phenomenon produced in the accretion flow around the
neutron star. It can, therefore, be used together with the highest
observed QPO frequency in placing upper bounds on the mass of
individual sources (Miller et al.\ 1998; see Fig.~5). This is a rather
model-independent argument and is applicable to almost all current
model of neutron-star variability. More specific comparison of particular
models to the data lead to more severe constraints on the properties of
neutron-star matter. For example, the observed correlations between QPO 
frequencies can be accounted for in the relativistic model only if the
neutron-star masses are $>1.8M_\odot$ and hence only if the equation of
state is relatively stiff.

Finally, the high frequencies of the observed QPOs strongly suggest
that they originate very close to the neutron-star surfaces or the
event horizons of black holes. As a result, they must carry signatures
of the strong gravitational fields in which they are produced, such as
the presence of an innermost stable circular orbit outside the compact
object or of a maximum in the radial epicyclic frequency. Evidence for
both phenomena have already been reported in the literature (Zhang et
al.\ 1998; Belloni et al.\ 2000; see also Fig.~5) and conclusive 
detection of them will provide us with the first direct evidence of
strong-field general relativistic effects.

\section*{ACKNOWLEDGEMENTS}

It is a pleasure to thank a number of people who contributed in many
different ways to the work presented in this article.  I would
especially like to thank T.\ Belloni, M.\ van der Klis and the
high-energy group at the Univ.\ of Amsterdam, as well as D.\
Chakrabarty, D.\ Galloway, M.\ Muno, and the MIT group on burst
oscillations. I am also grateful to F.\ \"Ozel for useful discussions
and carefully reading the manuscript and to M.\ Muno for help in
preparing several of the figures in this paper.

\end{document}